\documentclass[%
 reprint,
superscriptaddress,
 amsmath,amssymb,
 aps,
 prl
]{revtex4-2}
\usepackage{color}
\usepackage{graphicx}
\usepackage{dcolumn}
\usepackage{bm}
\usepackage[hidelinks]{hyperref}
\usepackage{lipsum}
\usepackage{float}
\makeatletter
\renewcommand{\l@section}[2]{\@dottedtocline{1}{0em}{2em}{#1}{#2}}
\makeatother

\begin{document}


\title{
Plasmonic Fourier Surfaces Revisited: \\
Relating Bandgaps with Bound States in the Continuum
}

\author{Alexander A. Antonov}
\email{alexander.antonov@tuhh.de}
\affiliation{Institute of Photonics, Hamburg University of Technology, 21073 Hamburg, Germany}
\affiliation{Chair in Hybrid Nanosystems, Nanoinstitute Munich, Faculty of Physics, Ludwig-Maximilians-Universität München, 80539 Munich, Germany}

\author{Connor Heimig}
\affiliation{Institute of Photonics, Hamburg University of Technology, 21073 Hamburg, Germany}
\affiliation{Chair in Hybrid Nanosystems, Nanoinstitute Munich, Faculty of Physics, Ludwig-Maximilians-Universität München, 80539 Munich, Germany}

\author{Hannah Niese}
\affiliation{Optical Materials Engineering Laboratory, Department of Mechanical and Process Engineering, ETH Zurich, 8092 Zurich, Switzerland}

\author{Yannik M. Glauser}
\affiliation{Optical Materials Engineering Laboratory, Department of Mechanical and Process Engineering, ETH Zurich, 8092 Zurich, Switzerland}

\author{Sander J. W. Vonk}
\affiliation{Optical Materials Engineering Laboratory, Department of Mechanical and Process Engineering, ETH Zurich, 8092 Zurich, Switzerland}

\author{David J. Norris}
\affiliation{Optical Materials Engineering Laboratory, Department of Mechanical and Process Engineering, ETH Zurich, 8092 Zurich, Switzerland}

\author{Maxim V. Gorkunov}
\email{gorkunov@crys.ras.ru}
\affiliation{National Research Nuclear University MEPhI Kashirskoe shosse 31, Moscow 115409, Russia}
\affiliation{Theoretical Physics and Quantum Technologies Department, National University of Science and Technology ‘MISIS’, Moscow 119049, Russia}

\author{Andreas Tittl}
\email{andreas.tittl@tuhh.de}
\affiliation{Institute of Photonics, Hamburg University of Technology, 21073 Hamburg, Germany}
\affiliation{Chair in Hybrid Nanosystems, Nanoinstitute Munich, Faculty of Physics, Ludwig-Maximilians-Universität München, 80539 Munich, Germany}


\begin{abstract}
Periodically corrugated metal interfaces supporting surface plasmon polaritons (SPPs) belong to the earliest nanoplasmonic platforms. Even simplest reliefs described by a few harmonics--plasmonic Fourier surfaces--display markedly different far-field signatures depending on the corrugation depth and symmetry: shallow reliefs exhibit a plasmonic bandgap (PBG) between two hybridized SPP standing waves, while deeper reliefs support asymmetry-induced single sharp resonances, termed over the last decade as quasi-bound states in the continuum (qBICs). Although these spectral features have long been observed experimentally and treated empirically, the underlying eigenmode evolution connecting the shallow and deep corrugation regimes has remained largely unexplored. Here we revisit this long-standing problem by analyzing it in terms  of modern eigenstate formalism. Starting from the Rayleigh hypothesis, we develop a concise first-principle analytical description that explicitly captures how one eigenmode transforms from a dark bound state into an observable qBIC, while the other turns from bright into an overcoupled, unobservable state--thus unifying the SPP manifestations featuring PBG and qBIC within the same eigenmode framework. Finally, we demonstrate the practical relevance of the theory by showing how precise eigenstate engineering can enhance the SPP refractive index sensitivity.
\end{abstract}

\maketitle

\makeatletter
\let\orig@addcontentsline\addcontentsline
\renewcommand{\addcontentsline}[3]{}
\makeatother


\phantomsection\pdfbookmark[1]{Introduction}{sec:intro}
\section*{Introduction}
Plasmons, also called plasmon-polaritons, arise from the coupling of collective oscillations of conduction electrons in metals with the electromagnetic field and attract vast attention due to the ability to confine light at subwavelength scales, overcoming the diffraction limit and providing exceptionally strong near-field enhancement~\cite{maier2007plasmonics}. They appear in various geometries: localized plasmons are hosted by metallic nanoparticles \cite{amendola2017surface} and single meta-atoms, while traveling surface plasmon polaritons (SPPs) are inherent to metal-dielectric interfaces \cite{maier2005plasmonics, zhang2012surface} and periodic structures--metamaterials and metasurfaces~\cite{meinzer2014plasmonic}. The periodicity of discrete meta-atom arrangements, as well as of continuous surface modulations, opens up additional degrees of freedom when co- and counter-propagating SPP waves become mixed into complex modes with tailored spectra and coupling to free-space light manifested by a variety of peculiar far-field resonant phenomena. 
These range from Rayleigh--Wood anomalies \cite{wood1902xlii,rayleigh1907}--the earliest studied diffraction phenomena directly linked to SPP excitation~\cite{hessel_new_1965}--and extraordinary optical transmission \cite{genet_light_2007,sturman2008theory,van2012quasi,rodrigo2016extraordinary}, to plasmonic bandgaps (PBGs) \cite{barnes1995photonic, barnes1996physical, kelf2005plasmonic} and asymmetry-controlled resonances originating from symmetry-protected bound states in the continuum (BICs) \cite{aigner2022plasmonic, wang2023mirror,seo2020fourier,li2025unidirectional}.

The periodic modulation can be fairly simple: a metal-dielectric interface with a profile described by just one Fourier harmonic already gives rise to a PBG, as it was predicted and observed in early studies of corrugated surfaces~\cite{barnes1995photonic, barnes1996physical, kelf2005plasmonic}. Multiple Bragg scattering shapes SPPs into two distinct standing-wave modes with different energy levels and a bandgap in-between~\cite{barnes1996physical}. Similarly to other photonic bandgaps, PBGs on surfaces profiled by one or a few Fourier harmonics have been widely utilized for steering and guiding of SPPs~\cite{bozhevolnyi2001waveguiding} and for controlling the coherent thermal emission \cite{biener2008highly}. Furthermore, by introducing selective patterning of high-index materials to create integrated cavities, one achieves localization of standing-wave  SPPs with the frequencies within the bandgap~\cite{kocabas2008plasmonic}. High local density of optical states at the band edges significantly enhances light-matter interaction, providing a powerful tool for tailoring the system's optical response and coupling regimes \cite{turnbull2001relationship, torma2014strong}.

Recent rapid growth of interest in resonant nanophotonics covers nanostructures based on metals, semiconductors and  dielectrics, which promise applications in diverse emerging photonic technologies~\cite{yao_integrated-resonant_2023}. Understanding the nature and contributions of the resonances to various phenomena has required the development of  appropriate theoretical concepts and tools from a relatively simple phenomenology of the coupled-mode theory (CMT) \cite{fan_temporal_2003} to elaborate resonant-state expansions~\cite{lalanne_light_2018}. The concept of photonic BICs happens to be especially fruitful for understanding and engineering high quality factor (Q-factor) resonances. A BIC is a specific eigenstate fully isolated from the continuum of free-space electromagnetic waves (despite sharing the same energy) through symmetry protection \cite{plotnik2011experimental} or destructive interference \cite{hsu2013observation} and is robust owing to topological protection~\cite{zhen2014topological}.
Its radiative counterpart, known as a quasi-BIC (qBIC), is formed by introducing subtle symmetry-breaking perturbations that precisely control its coupling to free-space light~\cite{koshelev2018asymmetric}. Widely exploited in dielectric photonics due to their extremely high Q-factors \cite{kodigala2017lasing,watanabe2025low}, qBICs enable strong light-matter interactions~\cite{ardizzone2022polariton,heimig2026chiral} and are desired for applications spanning from lasing \cite{kodigala2017lasing,hwang2022nanophotonic} and sensing \cite{maksimov2020refractive, wang2021all} to nonlinear \cite{koshelev2019nonlinear, liu2019high} and chiral \cite{gorkunov2020metasurfaces, kuhner2023unlocking} optics. 
In plasmonic systems, qBICs remain comparably less explored as ohmic losses inherent to metals limit the achievable Q-factors. Nevertheless, controlled radiation channels and topological protection \cite{shen2024topologically} endow plasmonic qBICs with compelling advantages desired for molecular sensing \cite{aigner2022plasmonic}, perfect absorbers \cite{wang2023mirror}, emission \cite{seo2020fourier}, and unidirectional coupling \cite{li2025unidirectional}. 

Being typically presented as distinct phenomena, PBGs and qBICs although manifest themselves differently in the far-field, share the same physical origin, as they arise due to the coupling of two otherwise independent SPP standing waves on a periodically corrugated metal surface, one of which is symmetry-protected and radiatively dark. This connection is not itself new. Barnes \textit{et al.} derived an analytic model for the PBG on shallow metallic gratings, obtaining explicit expressions for the central position and the gap width in terms of the symmetry-breaking corrugation amplitude \cite{barnes1995photonic,barnes1996physical}. Kreiter \textit{et al.} subsequently investigated deep and asymmetric gold gratings, classifying the resulting SPP resonances by the symmetry of their magnetic-field distribution, and showed experimentally that a symmetry-protected, non-radiative mode on a symmetric deep grating becomes observable once that symmetry is broken \cite{kreiter2002surface}. In other words, the authors demonstrated a particular realization of the qBIC concept which was generally formulated only a decade after that~\cite{hsu2013observation}. More recently, phenomenological CMTs based on a coupled-oscillator model have been used to describe similar modal hybridization in plasmonic-BIC systems of non-harmonic rectangular shapes~\cite{seo2020fourier, li2025unidirectional}, but with many model parameters obtained by fitting the results of numerical simulations with no explicit relations to the grating shape. However, these insights have existed only as mosaic pieces of a broader physical picture, primarily because analytical perturbation theory and phenomenological CMTs cover entirely different limiting cases of surface corrugation. What has remained missing, therefore, is a first-principles eigenstate theory that (i) reproduces both the shallow-grating PBG and the deep-grating qBIC regimes as limits of a single analytic model, (ii) derives the coupled-oscillator equations from first principles instead of phenomenologically presuming them, and (iii) connects all model parameters (mode frequencies, coupling strength, and decay rates) directly to the grating geometry and material parameters.

In this work, we address this gap by revisiting the long-standing problem of SPPs on periodically corrugated metal interfaces--so-called plasmonic Fourier surfaces \cite{lassaline2020optical}--using the apparatus of modern resonant nanophotonics. Such structures are presently experiencing a renewal of interest due to the development of nanofabrication techniques, including thermal scanning-probe \cite{lassaline2020optical,glauser2026fourier}, grayscale electron-beam \cite{shanks_metasurfaces_2025}, and focused ion beam \cite{kondratov2016extreme, gorkunov2018chiral} lithography, and holographic inscription \cite{lim2021field, oscurato2021large}. Fourier surfaces promise to become a versatile platform for multifunctional pixels capable of characterizing incoming light and generating arbitrary, background-free optical wavefronts \cite{glauser2026fourier}. Based on the Rayleigh hypothesis, we derive from first principles an explicit coupled-oscillator dispersion relation for the two SPP eigenmodes. This relation shows how, for shallow gratings, the asymmetry-induced hybridization of dark (plasmonic BICs) and bright standing waves opens the PBG, while for deep gratings, where the bright mode has already broadened into an unobservable background, the same hybridization instead transforms the dark mode into a single sharp qBIC. Motivated by the demand for plasmonic refractive-index sensing applications \cite{li2015plasmon,ye2014high, cai2019solution, li2023high}, we further demonstrate the capabilities of our eigenstate theory to significantly enhance the sensing performance of plasmonic Fourier surfaces.

\phantomsection\pdfbookmark[1]{Results}{sec:results}
\section*{Results}
\phantomsection\pdfbookmark[2]{General concept}{subsec:general-concept}
\subsection*{General concept}
\begin{figure}[b]
\includegraphics{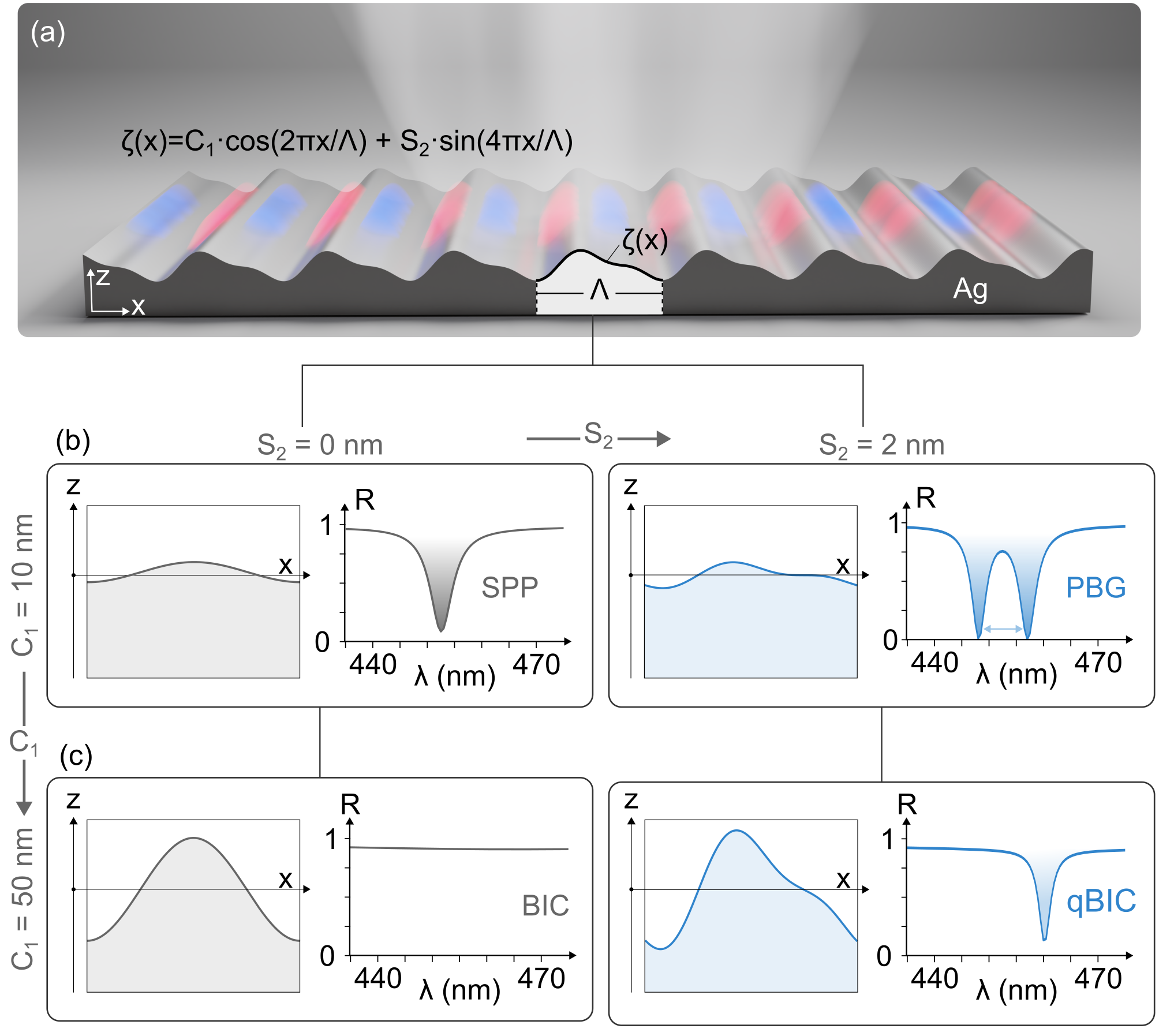}
\caption{Plasmonic Fourier surfaces exhibiting either PBG or qBIC spectra. 
(a)~Schematic of a silver Fourier surface  defined by the surface corrugation $\zeta(x)$ and excited by TM-polarized light. The grating period is $\Lambda = 420\,\mathrm{nm}$. 
(b)~Shallow corrugation regime defined by $C_1=10$~nm. For a symmetric corrugation ($S_2=0$), the reflectance spectrum exhibits a single SPP resonance. Introducing symmetry breaking via $S_2=2$~nm leads to the formation of a PBG.
(c)~Deep corrugation regime with $C_1=50$~nm. In the symmetric case ($S_2=0$), the structure supports a symmetry-protected BIC and therefore exhibits no resonant feature in reflectance. Breaking the symmetry by introducing $S_2=2$~nm transforms the BIC into a radiatively accessible qBIC.}\label{fig:fig1}
\end{figure}

We consider a one-dimensional plasmonic Fourier surface composed of silver with permittivity $\varepsilon_\mathrm{m}$ \cite{mcpeak2015plasmonic}, embedded in a dielectric environment with $\varepsilon_\mathrm{d}$. The periodic surface corrugation with a period (pitch) $\Lambda=420$~nm is defined by the function $\zeta(x)$ [see Fig.~\ref{fig:fig1}(a)], composed of the first two Fourier harmonics:
\begin{equation}\label{zeta}
    \zeta(x)=C_1\cos(2\pi x/\Lambda)+S_2\sin(4\pi x/\Lambda),
\end{equation}
where $C_1$ is the amplitude of the fundamental harmonic, while $S_2$ serves as the asymmetry parameter that breaks the relief parity with respect to the $x=0$ plane.

We first investigate a structure in vacuum ($\varepsilon_\mathrm{d}=1$) with a shallow symmetric corrugation defined solely by the fundamental harmonic with an amplitude $C_1=10$~nm. Using COMSOL Multiphysics based on the finite element method (FEM), we demonstrate in Fig.~\ref{fig:fig1}(b) the reflectance spectra (R) for a normally incident (against the $z$-axis) TM-polarized plane wave. The diffractive grating enables excitation of an SPP standing wave at a wavelength of $\lambda\approx455$~nm. Then, introducing a second harmonic amplitude $S_2=2$~nm breaks the symmetry of the corrugation, leading to the formation of a PBG \cite{barnes1996physical}. 
The resulting bandgap broadens with increasing $S_2$, as illustrated by the angle-resolved reflectance spectra shown in Fig.~\ref{fig:Bandgap} of the Supplementary Information (SI).

Next, we increase the amplitude of the fundamental harmonic $C_1$ and find that, for such a deep ($C_1=50$~nm) symmetric ($S_2=0$) grating, the SPP resonance can no longer be observed [see Fig.~\ref{fig:fig1}(c)]. However, breaking the symmetry by introducing a finite amplitude $S_2=2$~nm results in a pronounced resonance in the reflectance spectra. Moreover, as we show in Fig.~\ref{fig:SI_QvsS2}, the radiative quality factor ($Q_{\mathrm{rad}}$) of this resonance follows the general qBIC scaling rule $Q_{\mathrm{rad}}\propto S_2^{-2}$ where $S_2$ plays the role of the asymmetry parameter \cite{koshelev2018asymmetric}. These observations indicate that the deep grating supports a resonance with the characteristics of a qBIC originating from a symmetry-protected BIC. 

Therefore, such a simple one-dimensional diffraction grating exhibits two markedly different resonant regimes at different corrugation depths, historically explained separately: expressions for the PBG of shallow gratings were derived using a perturbative approach \cite{barnes1995photonic,barnes1996physical}, while the single sharp resonance of deep, asymmetric gratings was empirically demonstrated and explained within the symmetry-classification arguments \cite{kreiter2002surface}. Below we show, that both regimes emerge from the same pair of eigenmodes, whose hybridization we trace continuously from shallow to deeper gratings. The fundamental harmonic amplitude $C_1$ controls the transition between these regimes, while $S_2$ tunes the resonance properties through symmetry breaking, enabling the formation of PBGs and qBICs. We also note that all effects associated with a nonzero $S_2$ can also be achieved for a symmetric grating ($S_2=0$) away from the $\Gamma$-point (i.e., under oblique incidence), where a non-zero in-plane wave vector provides the required symmetry-breaking perturbation (see Fig.~\ref{fig:SI_R_vs_s2_and_phi}). In this work, however, we focus on the properties of the Fourier surfaces at the $\Gamma$-point, while noting that the results and analysis can be straightforwardly generalized over the reciprocal in-plane momentum space as well.

\phantomsection\pdfbookmark[2]{Analytical theory}{subsec:analytical-theory}
\subsection*{Analytical theory}
To address the question of how two distinct regimes coexist within a single system, and how they transit into each other, we analyze the underlying eigenstates. We employ the Rayleigh hypothesis (RH) to expand the eigenmode fields of the Fourier surface as sums of outgoing and evanescent TM-polarized waves \cite{rayleigh1907dynamical}. Accordingly, the $y$-component of the magnetic field amplitude is written as:
\begin{equation} \label{eq:Hy-RH}
    H_y(x,z)=
    \begin{cases}
\sum_u a_ue^{ip_ux-ik_uz},\ z<\zeta(x) \\
\sum_u b_ue^{ip_ux+i\varkappa_uz},\ z>\zeta(x)
\end{cases}
\end{equation}
where monochromatic time dependence $e^{-i\omega t}$ is assumed and the field amplitudes in dielectric and metal are $a_u$ and $b_u$ respectively. The corresponding wavenumbers are presented in the form:
\begin{equation}\label{eq:wavevectors}
\begin{gathered}
     k_u=\sqrt{\varepsilon_\mathrm{d}\left(\frac{2\pi}{\lambda}\right)^2-p^2_u}, \ \ \ \varkappa_u=\sqrt{\varepsilon_\mathrm{m}\left(\frac{2\pi}{\lambda}\right)^2-p^2_u},
\end{gathered}
\end{equation}
where $p=2\pi/\Lambda$ is the lattice wavenumber, $p_u = pu$, and $u \in \mathbb{Z}$ denotes the diffraction order.

 The RH has established limits of  applicability only for simple interfaces, such as single harmonic \cite{millar1969rayleigh}, triangular, or rectangular \cite{van1979rayleigh} gratings. For more complex surface profiles, its validity must be determined empirically, for example, by comparison with full scale numerical simulations. It has also been noted that the separation of the near fields into incident and reflected waves is not entirely unambiguous; nevertheless, one may expect quantitative accuracy and a wide range of applicability from this seemingly rough approximation as described in Ref.~\cite{voronovich2007rayleigh}. 

In addition to the RH, other approaches can be employed, such as the Chandezon method, which involves a coordinate transformation flattening the corrugated surface \cite{chandezon1982multicoated}. Alternatively, the system can be decomposed into several homogeneous layers, whose transfer matrices can be combined to construct the scattering matrix of the entire system \cite{tikhodeev2002quasiguided}. Although other analytical approaches may provide higher quantitative accuracy, we demonstrate below that the RH, in its simplest representation accounting for only three field harmonics, yields a simple analytical expression that qualitatively captures the eigenstates and, consequently, explains the transition between distinct modal regimes.

To obtain the characteristic eigenstate dispersion equation, we apply the boundary conditions at the corrugated surface $z=\zeta(x)$, which require the continuity of $H_y$ and of its derivative along the local surface normal divided by the permittivity $\varepsilon^{-1}\partial H_y/\partial n$. Then, following the procedure described in \cite{tishchenko2009numerical,antonov2019corrugated,antonov2021dielectric}, we multiply the continuity equations by $e^{-ip_vx}$, where $v\in \mathbb{Z}$, and integrate by parts over one period $\Lambda$. 
This yields two sets of linear equations for the field amplitudes $a_u$ and $b_u$: 
\begin{equation} \label{set_of_rh_eq}
\begin{gathered}
 \sum_{u}a_{u}I_{vu}  = \sum_{u} b_{u}J_{vu}\\
	-\sum_{u}a_{u}I_{vu}\frac{k_{0}^{2}-p_{u}p_{v}}{\varepsilon_\mathrm{d}k_{u}}  =
		\sum_{u}\frac{\varkappa_{0}^{2}-p_{u}p_{v}}{\varepsilon_\mathrm{m}\varkappa_{u}}b_{u}J_{vu},
        \end{gathered}
\end{equation}
where the integrals $I_{vu}$ and $J_{vu}$ depend on the corrugation $\zeta(x)$ and are indexed by the integers $v$ and $u$:
\begin{equation}\label{eq:integrals}
\begin{gathered}
I_{vu}=\frac{1}{\Lambda}\int_{0}^{\Lambda}\exp\left[i(u-v)\frac{2\pi}{\Lambda}x- ik_{u}\zeta(x)\right]dx,\\
	J_{vu}=\frac{1}{\Lambda}\int_{0}^{\Lambda}\exp\left[i(u-v)\frac{2\pi}{\Lambda}x+ i\varkappa_{u}\zeta(x)\right]dx.
    \end{gathered}
\end{equation}
As discussed in \cite{tishchenko2009numerical,antonov2019corrugated}, the number of field harmonics $2N+1$ $(|v|\le N,|u|\le N)$ retained in Eq.~\eqref{set_of_rh_eq} represents a trade-off.  On the one hand, increasing $N$ improves the accuracy and allows capturing some minor features. On the other hand, excessively large $N$ requires integration of rapidly oscillating functions, leading to the accumulation of numerical errors and, ultimately, to nonphysical results. To determine an appropriate value of $N$ we benchmark our approach by solving the reflectance problem for Fourier surfaces with both shallow ($C_1=10$~nm) and deep ($C_1=50$~nm) corrugations. As shown in Fig.~\ref{fig:SI_RH_with_dif_Q}, retaining only the $-1^{\mathrm{st}}$, $0^{\mathrm{th}}$ and $+1^{\mathrm{st}}$ field harmonics $(N=1)$ is sufficient to qualitatively reproduce all necessary resonance features, such as number of the resonances, their spectral positions, and linewidths. Including the next field harmonics $(N=2)$ yields reflectance spectra indistinguishable from the FEM results. However, to avoid unnecessary mathematical complexity, we restrict ourselves to the minimal number of field harmonics with $N=1$. We also note that the only non-evanescent harmonic corresponds to $a_0$: the fields associated with $a_{\pm1}$ exponentially decay with $z$ because the eigenstate wavelengths lie beyond the diffraction threshold, while all harmonics with coefficients $b_u$ decay into metal [see the sketch in Fig.~2(a)].

\begin{figure}[h]
\includegraphics{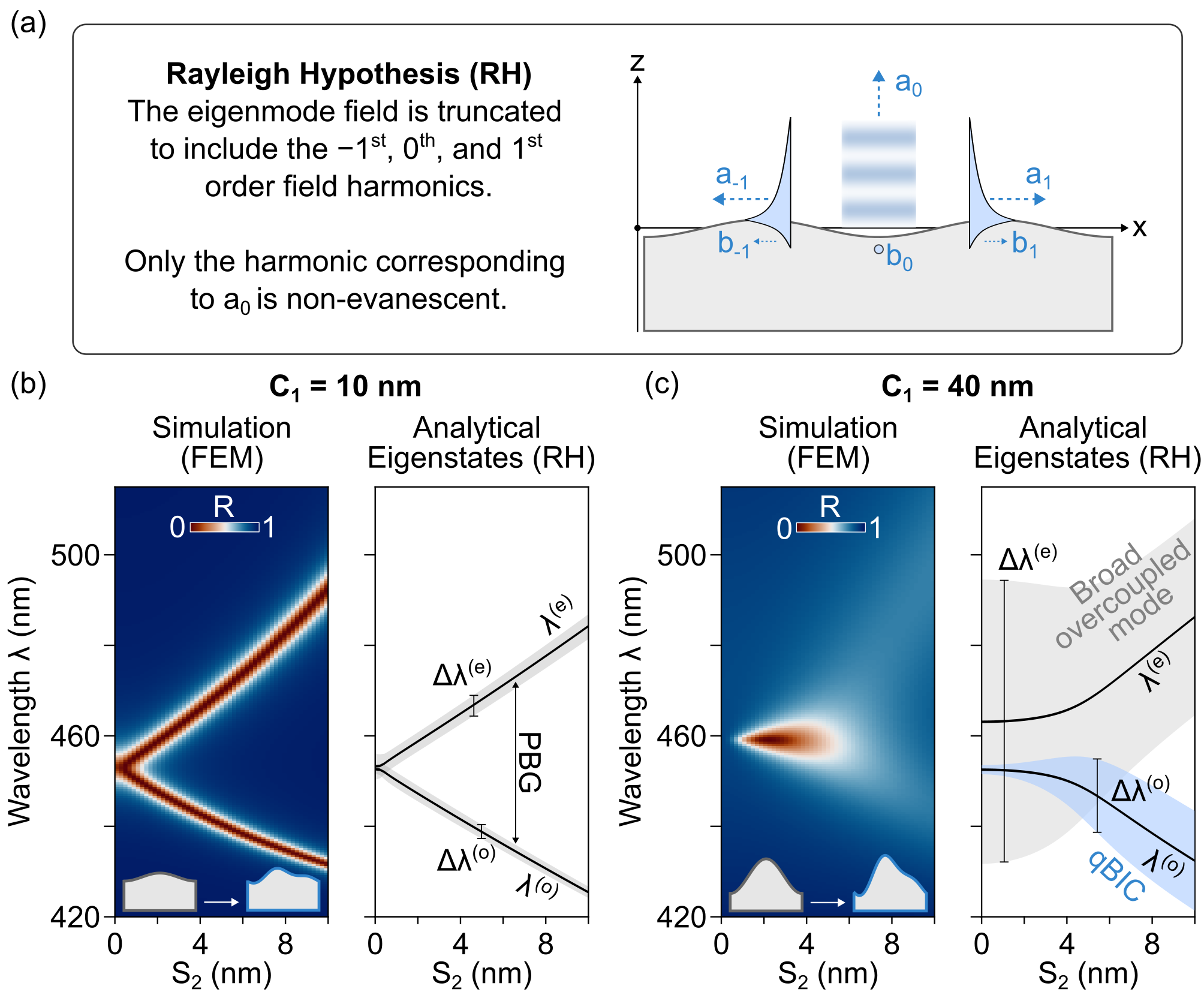}
\caption{\label{fig:fig2} 
(a)~Schematic illustration of the problem within the RH, where the fields of the eigenmodes inside a metal with surface corrugation $\zeta(x)$ are approximated using only the $-1^{\mathrm{st}}$, $0^{\mathrm{th}}$ and $+1^{\mathrm{st}}$ field harmonics. Dashed arrows indicate the directions in which the wavevector components of field harmonics are real. Only the field harmonic associated with $a_0$ is propagating (non-evanescent), whereas all other harmonics decay exponentially with $z$.
(b)~Comparison between FEM-based reflection spectra and analytical eigenstates within the RH framework (namely their spectral positions and linewidths) as functions of the perturbation amplitude $S_2$ for a Fourier surface with $\varepsilon_\mathrm{m}=-7.2+0.2i$, shallow corrugation ($C_1 = 10\,\mathrm{nm}$) and period $\Lambda=420$~nm. A PBG is manifested in the spectra for nonzero $S_2$.
(c)~Same comparison for a deeper corrugation ($C_1 = 40\,\mathrm{nm}$). In the symmetric case ($S_2=0$), the structure supports a symmetry-protected BIC together with another mode possessing a linewidth too broad to be spectrally resolved. Introducing nonzero $S_2$ breaks the symmetry, resulting in a qBIC with a finite linewidth $\Delta\lambda^{(\mathrm{o})}$, while the other mode remains too broad and overcoupled. Superscripts indicate mode parity at $S_2=0$: even (e) or odd (o).
}
\end{figure}

As we discuss in more detail in SI Note~\ref{SI:Note1}, the system of equations~\eqref{set_of_rh_eq} allows one to write the dispersion equation based on $3\times3$ matrices as we retain only the first three field harmonics ($N=1$). The matrix elements depend on the integrals~\eqref{eq:integrals}, which we expand in the limit of a shallow corrugation where both $C_1k_0\ll1$ and $S_2k_0\ll1$. Notably, the surface relief $\zeta(x)$ (see Eq.~1) determines that the eigenmode spectrum should be an even function of both $C_1$ and $S_2$. Indeed, inverting the sign of $C_1$ is equivalent to shifting the coordinate origin by $\Lambda/2$, which cannot alter the eigenmode spectrum. Similarly, inverting the sign of $S_2$ corresponds to a reversal of the $x$-axis direction, which also leaves the spectrum invariant. This, in particular, requires performing all expansions in powers of $C_1$ and $S_2$ up to the second order (see SI Note~\ref{SI:Note1}).

First, we consider the zeroth-order approximation with $C_1=0$ and $S_2=0$, and obtain the corresponding dispersion equation for a flat metal surface as:
\begin{equation}\label{eq:0th order}
    (\varepsilon_\mathrm{d}+\sqrt{\varepsilon_\mathrm{m}})(\varepsilon_\mathrm{d}\varkappa_1+\varepsilon_\mathrm{m} k_1 )^2=0,
\end{equation}
where the former factor never turns to zero as the dielectric with $\varepsilon_\mathrm{d}$ is assumed to be lossless. The second factor yields two identical eigenfrequencies:
\begin{equation}\label{eq:omega0}     \omega_0=pc\sqrt{\frac{\varepsilon_\mathrm{d}+\varepsilon_\mathrm{m}}{\varepsilon_\mathrm{m}\varepsilon_\mathrm{d}}},
\end{equation}
where $c$ is the speed of light.
Thus, a flat metal surface supports two degenerate eigenmodes with equel frequencies \eqref{eq:omega0}, both corresponding to a classical surface plasmon solution with a wavenumber equal to the lattice wavenumber $p$.

Next, we consider the first-order approximation, where the eigenfrequencies $\omega_\zeta$ of the corrugated surface experience a small detuning $\Delta\omega$ from the flat-metal eigenfrequency $\omega_0$:
\begin{equation}
\omega_\zeta\approx\omega_0+\Delta\omega.   
\end{equation}
Expanding all wavenumbers $k_u$ and $\varkappa_u$ in the vicinity of their values at $\omega_0$ and solving the corresponding dispersion equation yields:
\begin{multline}\label{eq:deltaequation}
        (\Delta\omega)^2 + pc\frac{(pC_1)^2}{2} \frac{\sqrt{\varepsilon_\mathrm{d}}+i\sqrt{-\varepsilon_\mathrm{m}}} {W_\mathrm{m}\varepsilon_\mathrm{d}-W_\mathrm{d}\varepsilon_\mathrm{m}}  \Delta\omega\\+ (pc)^2(pS_2)^2\frac{(\varepsilon_\mathrm{d}-\varepsilon_\mathrm{m})^2}{(\varepsilon_\mathrm{d} + \varepsilon_\mathrm{m})(W_\mathrm{m}\varepsilon_\mathrm{d}-W_\mathrm{d}\varepsilon_\mathrm{m})^2}=0,
\end{multline}
where 
\begin{equation}
    W_j=1+\frac{\omega_0}{2\varepsilon_j(\omega_0)}\frac{d\varepsilon_j}{d\omega}(\omega_0), \ \ \ j=\mathrm{m},\mathrm{d}.
\end{equation}
accounts for the material dispersion. It can be important, for example, for a Drude-type metal permittivity dispersion, when the factor $W_\mathrm{m}$ tends to vanish 
at large frequencies.
Note that the dispersion equation~\eqref{eq:deltaequation} retains the term proportional to $C_1^2\Delta\omega$ and not to $C_1^2$, whereas $S_2$ enters only via the $S_2^2$ term, highlighting different roles of the two harmonic corrugation amplitudes.

For better clarity, we substitute $\Delta\omega=\omega_\zeta-\omega_0$ back into the dispersion equation~\eqref{eq:deltaequation} and rewrite it in the coupled-oscillators form:
\begin{equation}\label{eq:omegacs_disp}
    (\omega_{\zeta}-\omega_0)(\omega_{\zeta}-\omega_C)=g^2,
\end{equation}
where the eigenfrequency $\omega_C$ is determined solely by the fundamental cosine surface harmonic: 
\begin{equation} \label{eq:omega_C}
    \omega_C=\omega_0-pc\frac{(pC_1)^2}{2}\frac{\sqrt{\varepsilon_\mathrm{d}}+i\sqrt{-\varepsilon_\mathrm{m}}} {W_\mathrm{m}\varepsilon_\mathrm{d}-W_\mathrm{d}\varepsilon_\mathrm{m}}, 
\end{equation}
and the parameter $g$ defined as:
\begin{equation}\label{eq:u}
    g=pc(pS_2)\frac{\varepsilon_\mathrm{d}-\varepsilon_\mathrm{m}}{(W_\mathrm{m}\varepsilon_\mathrm{d}-W_\mathrm{d}\varepsilon_\mathrm{m})\sqrt{-(\varepsilon_\mathrm{m}+\varepsilon_\mathrm{d})}}.
\end{equation}
quantifies the coupling between two otherwise independent eigenmodes induced by a nonzero $S_2$. These modes have a clear meaning in the absence of $S_2$: the mode with $\omega_C$ is even with respect to the $x=0$ plane and couples to free space, whereas the other mode is odd and remains nonradiative (a BIC) with the frequency equal to that of a flat-metal plasmon $\omega_0$. Note that similar symmetry arguments for the symmetric grating were established in Refs.~\cite{barnes1995photonic,kreiter2002surface}. 

In the presence of $S_2$ perturbation, the two solution branches of Eq.~\eqref{eq:omegacs_disp}, given by:
\begin{equation}\label{eq:omega_pm}
    \omega_\zeta=\frac{1}{2}(\omega_0+\omega_C)\pm\sqrt{\frac{1}{4}(\omega_0-\omega_C)^2+g^2},
\end{equation}
correspond to hybrids built from initially odd and even modes. Both hybrid modes are  coupled to free-space waves, and their eigenfrequencies are complex even in the absence of absorption in metal, as evidenced by the nonzero imaginary part of $\omega_C$ according to Eq.~\eqref{eq:omega_C}. Therefore, for a nonzero $S_2$, the initially dark odd mode (BIC) acquires radiative coupling through the bright mode and transforms into a qBIC.

To summarize this section, the eigenmodes of the asymmetric grating originate from two degenerate standing waves (even and odd) on the flat-metal surface with the eigenfrequency $\omega_0$ given by Eq.~\eqref{eq:omega0}. The introduction of the fundamental corrugation amplitude $C_1$ results in radiative coupling of the even mode, while the odd mode remains a nonradiative BIC pinned at $\omega_0$. Finally, the perturbation $S_2$ induces hybridization of the modes, described by the coupled-oscillator dispersion equation~\eqref{eq:omegacs_disp}: both new hybrid solutions become coupled to free-space radiation, with the odd mode acquiring a finite radiative linewidth typical of a qBIC. To the best of our knowledge, this is the first time such a coupled-oscillator model for the hybridization of different-parity plasmonic modes has been obtained explicitly and from first principles, rather than introduced phenomenologically \cite{seo2020fourier, li2025unidirectional}.

\phantomsection\pdfbookmark[2]{Eigenstates analysis}{subsec:eigenstates-analysis}
\subsection*{Eigenstates analysis}

These analytical results now allow us to investigate the continuous transition between the PBG regime of shallow corrugations to the single sharp qBIC resonance of deep corrugations. First, we note that according to Eq.~\eqref{eq:deltaequation}, the frequency dispersion of both metal and dielectric permittivities effectively renormalizes the frequency shift as $(W_\mathrm{m}\varepsilon_\mathrm{d}-W_\mathrm{d}\varepsilon_\mathrm{m})(\varepsilon_\mathrm{d}-\varepsilon_\mathrm{m})^{-1}\Delta\omega\rightarrow \Delta\omega$ and does not cause any drastic consequences. 
Therefore, in this section, we assume that the Fourier surface is in vacuum ($\varepsilon_\mathrm{d}=1$) and we fix the silver permittivity at $\varepsilon_\mathrm{m}=-7.2+0.2i$, consistent with tabulated data at a wavelength of $450$~nm \cite{mcpeak2015plasmonic}, corresponding to the eigenfrequency Eq.~\eqref{eq:omega0} for the chosen period of $\Lambda=420$~nm. 

In Figs.~2(b) and (c) we plot the reflectance spectra of the Fourier surface as a function of the perturbation $S_2$ obtained using FEM simulations for shallow $C_1=10$~nm and deep $C_1=40$~nm corrugations, respectively. Along with the simulations, we also plot both the resonance positions
$\lambda$=\ $2\pi c /\mathrm{Re}[\omega_{\zeta}]$ and linewidths (FWHM) $\Delta\lambda$=\ $4\pi c \ \mathrm{Im}[\omega_\zeta]/\mathrm{Re}[\omega_\zeta]^2$ of the analytical eigenstates in accordance with Eq.~\eqref{eq:omegacs_disp}.
Superscripts $\mathrm{e}$ and $\mathrm{o}$ denote the parity of the mode at $S_2=0$: either the even bright mode [minus sign in front of the square-root term in Eq.~\eqref{eq:omega_pm}], or the odd dark (BIC) mode [plus sign in Eq.~\eqref{eq:omega_pm}].

In the case of shallow ($C_1=10$~nm) and symmetric ($S_2=0$) surface corrugation [Fig.~\ref{fig:fig2}(b)], only the even mode couples to the far field, while the odd mode remains nonradiative and pinned to the surface plasmon frequency $\omega_0$, corresponding to the wavelength of $\lambda\approx455$~nm. 
Introducing a nonzero $S_2$ perturbation induces hybridization between the modes, resulting in two distinct SPP resonance branches separated by a PBG. As follows directly from Eq.~\eqref{eq:omegacs_disp}, increasing $S_2$ widens the PBG, as the coupling parameter scales linearly with the perturbation $g\propto S_2$ [Eq.~\eqref{eq:u}]. Regarding the linewidths, the resonance at longer wavelengths broadens, while the one at shorter wavelengths narrows as $S_2$ increases. Moreover, the decay rates of both hybrid modes remain close to the critical coupling condition (equal contributions from radiative and dissipative losses), resulting in resonant dips clearly resolved in reflection spectra. All these trends are consistently captured by both the numerical simulations and the analytical theory.

For the deeply corrugated Fourier surface with $C_1=40$~nm and $S_2=0$ [Fig.~\ref{fig:fig2}(c)], the odd mode remains nonradiative (i.e., a BIC), while the even mode's radiative losses grow too high above dissipative losses thus dragging the mode too far from the critical coupling condition and leaving it overcoupled and too broad to be resolved in the reflection spectra. The nonzero linewidth $\Delta\lambda^{(\mathrm{o})}$ at $S_2=0$ [Fig.~\ref{fig:fig2}(c)] originates solely from dissipation losses and does not correspond to an observable spectral feature. We also note that, in contrast to the results of \cite{barnes1995photonic,barnes1996physical}, our expression for the PBG width [the doubled square-root term of Eq.~\eqref{eq:omega_pm}] depends not only on $S_2$, but also on $C_1$, and both contribute to the expression under the square root with different powers: as $S_2^2$ and $C_1^4$. Consequently, a finite frequency separation between the two eigenmodes already exists at $S_2=0$, as seen in Fig.~2(c). Introducing the perturbation $S_2$ then leads to mode hybridization, causing the BIC to evolve into a qBIC. Its radiative Q-factor follows the scaling $Q_{\mathrm{rad}}\propto S_2^{-2}$ (see Fig.~\ref{fig:SI_QvsS2}) accompanied by a blueshift in resonance position while the loss distribution remains close to the critical coupling condition. Meanwhile, the linewidth of the broad mode slightly narrows, allowing a faint trace to emerge in the reflectance spectra as $S_2$ increases. However, with a further increase in $C_1$, this trace becomes barely observable [see Fig.~\ref{fig:SI_R_vs_s2_and_phi}(c)]. 
Although $C_1=40$~nm may already push the limits of our approximation, the analytical theory still qualitatively-well captures all the main resonant features and explains the transition between shallow and deep gratings regimes.

\begin{figure}[t]
\includegraphics{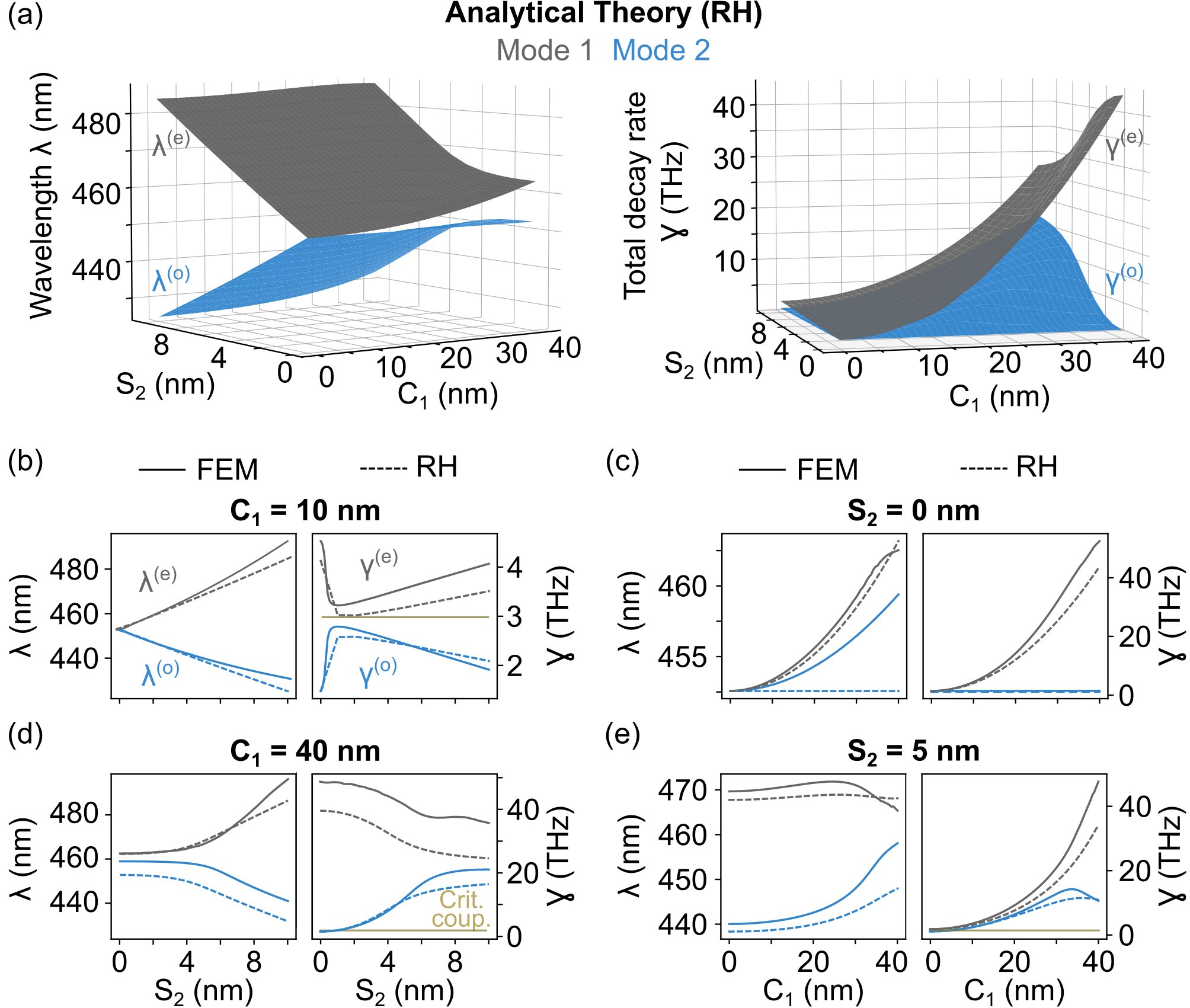}
\caption{\label{fig:fig3}
Eigenfrequency analysis of a Fourier surface with $\Lambda=420$~nm and $\varepsilon_\mathrm{m}=-7.2+0.2i$.
(a) Resonant wavelength $\lambda$ (left) and total decay rates $\gamma$ (right) as functions of the relief amplitudes $C_1$ and $S_2$ for the two hybrid modes, obtained from Eq.~\eqref{eq:omega_pm}. (b) Separate cuts at a fixed $C_1 = 10\,\mathrm{nm}$ (shallow corrugation) and $C_1 = 40\,\mathrm{nm}$ (deep corrugation) comparing the theory (dashed lines) and simulations (solid lines) for both $\lambda$ and $\gamma$ as functions of $S_2$.
(c) Analogous cuts at a fixed $S_2 = 0$ and $S_2 = 5\,\mathrm{nm}$ showing $\lambda$ and $\gamma$ as functions of $C_1$.
The yellow line in (b), (d) and (e)  marks the critical coupling condition ($\gamma_{\mathrm{dis}}=\gamma_{\mathrm{rad}}$) with $\gamma=\gamma_{\mathrm{dis}}+\gamma_{\mathrm{rad}}=3$~THz. Superscripts indicate mode parity at $S_2=0$: even (e) or odd (o).}
\end{figure}

Next, we examine in more detail how the corrugation amplitudes $C_1$ and $S_2$ affect the eigenmodes. Fig.~\ref{fig:fig3}(a) shows the resonant wavelengths $\lambda$ and the total decay rate $\gamma=\mathrm{Im}[\omega_\zeta]/(2\pi)$ (including both dissipative $\gamma_{\mathrm{dis}}$ and radiative $\gamma_{\mathrm{rad}}$ contributions) as functions of $C_1$ and $S_2$ for the hybrid eigenstates given by Eq.~\eqref{eq:omega_pm}. 
For clearer visualization, we plot representative cuts at fixed $C_1$ [Figs.~\ref{fig:fig3}(b) and (d)] and $S_2$ [Figs.~\ref{fig:fig3}(c) and (e)] and compare them with the FEM results.

For shallow corrugation with $C_1=10$~nm, the total decay rates $\gamma$ of both modes remain close to the critical coupling condition across the presented $S_2$ range: the dissipative contribution $\gamma_\mathrm{dis}=1.5$~THz is set by the material losses, so critical coupling ($\gamma_\mathrm{dis}=\gamma_{\mathrm{rad}}$) is reached at a total decay rate $\gamma=\gamma_\mathrm{dis}+\gamma_\mathrm{rad}=3$~THz, giving rise to a clearly resolved PBG. Further increase of $S_2$ results in a characteristic  "scissors"-like pattern for both the resonance position and the decay rate: $\lambda^{(\mathrm{e})}$ and $\gamma^{(\mathrm{e})}$ of the initially bright even mode increase, whereas $\lambda^{(\mathrm{o})}$ and $\gamma^{(\mathrm{o})}$ decrease. The latter observation is of particular interest: by introducing $S_2$ one can reduce the total decay rate of the system, mitigating dissipation effects and increasing the overall Q-factor. Such resonance line-narrowing is highly beneficial for applications requiring high spectral resolution, most notably refractive index sensing, where a smaller FWHM directly improves the sensing performance, as demonstrated in the following section.

In the cuts at $C_1=40$~nm shown in Figs.~\ref{fig:fig3}(d), the decay rates follow the trend discussed for Fig.~\ref{fig:fig2}(c): at $S_2=0$ the odd mode remains nonradiative (BIC), while $\gamma^{(\mathrm{e})}$ of the even mode is too large, so the mode is overcoupled and acts as a ``flat''background in the reflectance spectra. Introducing $S_2$ hybridizes the modes: $\gamma^{(\mathrm{o})}$ of the qBIC is drawn toward the critical coupling condition, giving a single sharp far-field resonance, while $\gamma^{(\mathrm{e})}$ remains too large.

Cuts at fixed $S_2$ [Figs.~\ref{fig:fig3}(c) and (e)] instead reveal how the eigenstates depend on $C_1$. For symmetric structures with $S_2=0$, the FEM results predict redshifts for both modes as $C_1$ increases. The analytical theory, however, captures only the redshift of the even mode, while the odd mode (BIC) remains pinned to $\omega_0$, in accordance with Eq.~\eqref{eq:omega_pm}. Nevertheless, the theory reliably describes the dependence of the decay rates $\gamma$ on $C_1$, which is particularly important for characterizing the transition between the shallow and deep corrugation regimes. Moreover, it also  reproduces qualitatively well the behavior of the hybrid states at $S_2=5$~nm, both in terms of  the resonance wavelength and the total decay rate.
\begin{figure}[b]
\includegraphics{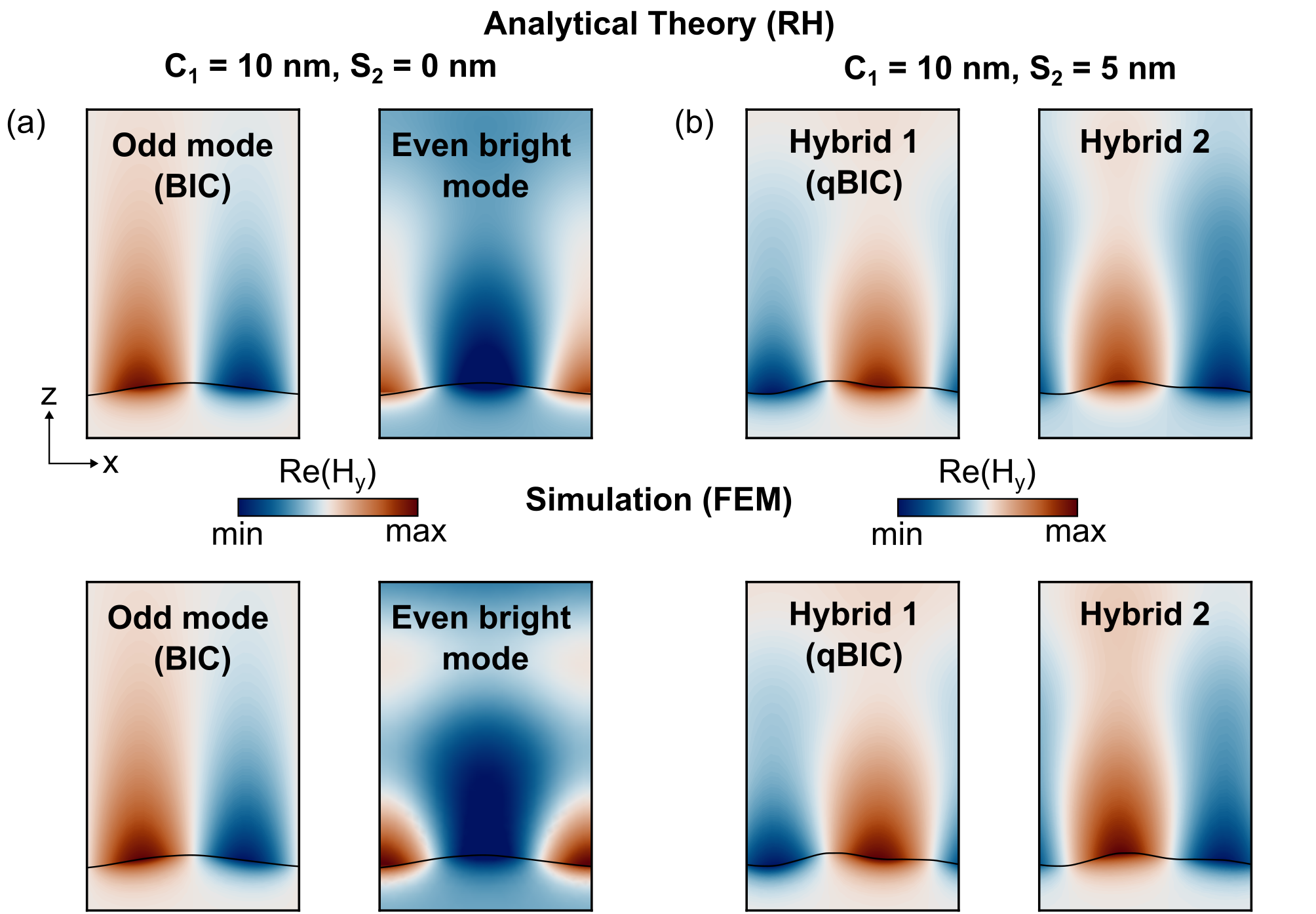}
\caption{\label{fig:fig4} 
Real part of the eigenmode fields $H_y(x,z)$ obtained using RH (top) and FEM (bottom) for (a) symmetric ($S_2=0$) and (b) asymmetric ($S_2=5$~nm) Fourier surfaces with a shallow corrugation ($C_1=10$~nm), period $\Lambda=420$~nm and $\varepsilon_\mathrm{m}=-7.2+0.2i$.
Fields are shown over one period $\Lambda$ along $x$; the black curve denotes the metal surface.
}
\end{figure}

Knowing the eigenfrequencies Eq.~\eqref{eq:omega_pm} allows us to determine the eigenvectors $a_u$ and $b_u$ and subsequently reconstruct the eigenmode fields $H_y(x,z)$ in accordance with Eq.~\eqref{eq:Hy-RH} for a more intuitive and clear visualization. In Fig.~\ref{fig:fig4} we plot the spatial distribution of the real part of $H_y$ for Fourier surfaces with a shallow relief ($C_1=10$~nm). Since the eigenmode fields are determined up to an arbitrary phase, we normalize it, such that the magnetic field is purely real at its nodes (points of maximum $|H_y|$). The simple analytical solution \eqref{eq:omega_pm} reproduces FEM results both qualitatively and quantitatively. 
The symmetric structure with $S_2=0$ (Fig~\ref{fig:fig4}(a)) indeed supports odd (BIC) and even (bright) eigenmodes with respect to the $x=0$ plane. In contrast, the structure with $S_2=5$~nm [Fig.~4(b)] lacks the mirror symmetry that would prevent mode coupling, resulting in the formation of two hybrid modes with finite linewidths and an associated spectral PBG. Analogous field distributions for Fourier surfaces that spectrally exhibit a single qBIC with $C_1=40$~nm are shown in Fig.~\ref{fig:SI_Eigenfields_C1=40nm}. Although the fundamental amplitude $C_1$ is relatively large for the present analytical treatment, the theory still qualitatively well captures the main eigenmode field features. 

Thus, the main conclusions regarding the eigenmodes are as follows. A Fourier surface with a shallow, symmetric ($S_2=0$) corrugation supports a bright SPP resonance (even mode) with a relatively small radiative decay rate, allowing its excitation at the $\Gamma$-point, while the odd mode remains a nonradiative BIC. Introducing asymmetry transforms the BIC into a qBIC, giving both hybrid modes comparable decay rates close to the critical coupling condition and yielding two pronounced resonances separated by a PBG. As the fundamental relief amplitude $C_1$
increases at $S_2=0$, the even mode's decay rate grows nonlinearly until the resonance is hardly observable in the far field, while the odd mode remains pinned and nonradiative. Introducing $S_2$
in this deep-corrugation transforms the BIC into a qBIC without significantly reducing the decay rate of the already-overcoupled even mode, which instead forms a flat background for the sharp qBIC.

\phantomsection\pdfbookmark[2]{Application to refractive index sensing}{subsec:ri-sensing}
\subsection*{Application to refractive index sensing}
As briefly mentioned in the previous section, increasing $S_2$ reduces the total decay rate $\gamma^{(\mathrm{o})}$ of the qBIC originating from the odd mode of a shallow relief ($C_1=10$~nm), which can help mitigating the effects of material dissipation. In this section, we demonstrate how this general prediction of the analytical theory can be utilized to significantly enhance the refractive index sensing performance--the most common and widely utilized application of plasmonic resonances \cite{li2015plasmon}.

\begin{figure}[h]
\includegraphics[width=1\linewidth]{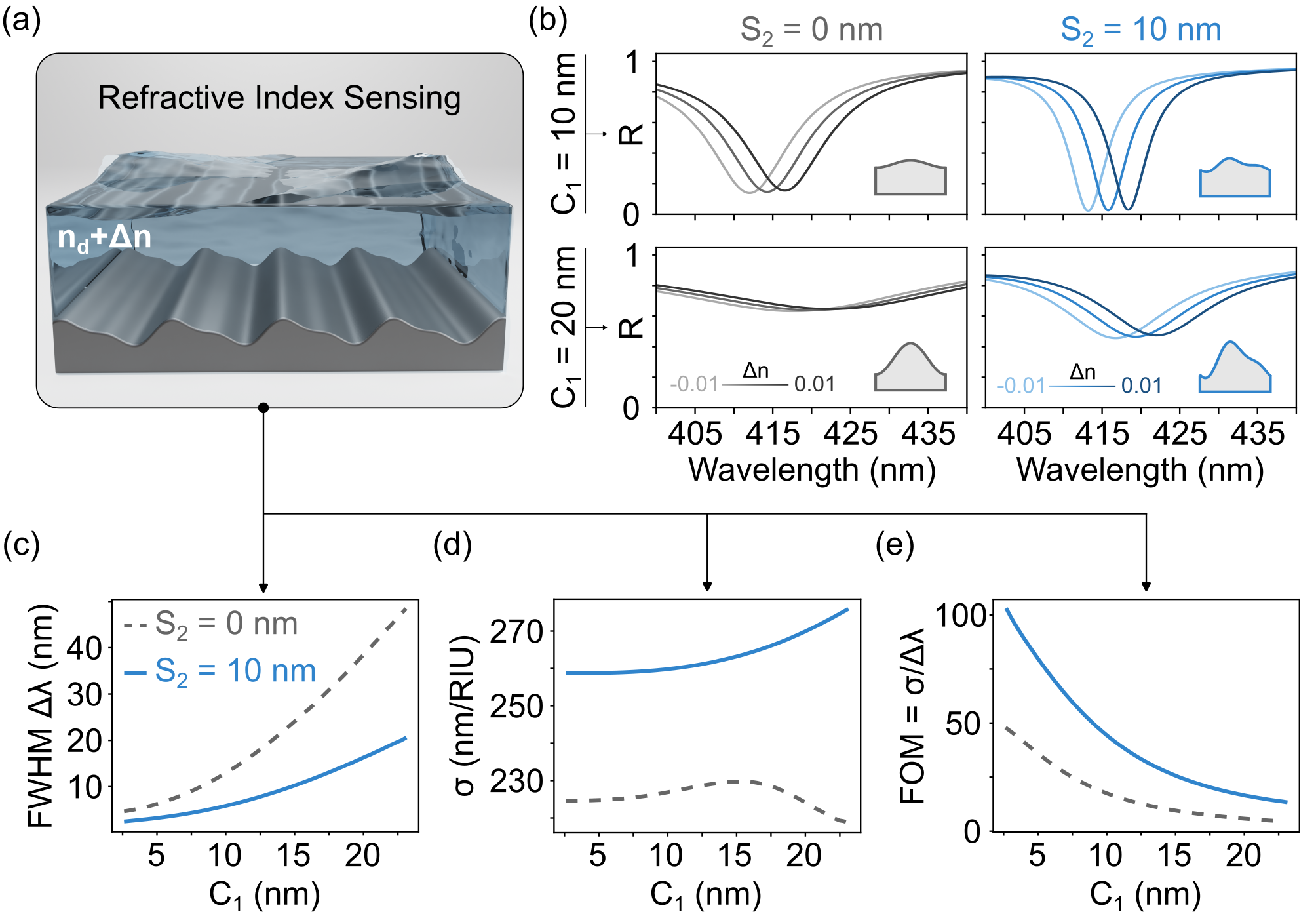}
\caption{\label{fig:fig5} 
(a)~Refractive index sensing in a water-like environment with $n_\mathrm{d}=1.33$ using silver Fourier surfaces. (b)~Reflectance spectra of symmetric ($S_2=0$) and asymmetric ($S_2=10$~nm) structures in an environment with $n_\mathrm{d}+\Delta n$. Spectra are plotted for Fourier surfaces with the main amplitudes $C_1=10$~nm and $C_1=20$~nm. 
In the latter case, the differences between the spectra for different $\Delta n$ of the symmetric structure are barely distinguishable. (c)~FWHM ($\Delta\lambda$), (d)~sensitivity $\sigma$ and (e)~figure of merit ($\sigma/\Delta\lambda$) as functions of the main amplitude $C_1$ for symmetric and asymmetric Fourier surfaces. Structures with $S_2=10$~nm exhibit a more than two times lower FWHM and, therefore, better sensing performance.
}
\end{figure}

We consider a water-like environment surrounding the structure [see Fig.~\ref{fig:fig5}(a)], characterized by a refractive index of $n_\mathrm{d}=1.33$ and $\varepsilon_\mathrm{d}=n_\mathrm{d}^2$. We aim to track small changes in $n_\mathrm{d}$ by monitoring the shift of the resonance reflection dip, whose FWHM naturally limits the sensing accuracy. We use tabulated data for silver \cite{mcpeak2015plasmonic} and renormalize the Fourier surface period so that the new $\Lambda=290$~nm yields the same plasmon frequency $\omega_0$ in the aqueous environment as in the previous sections, corresponding to a wavelength of $455$~nm. First, we consider a surface with a shallow corrugation with $C_1=10$~nm. By introducing a second Fourier harmonic with $S_2$, two resonances separated by a PBG appear, and we track the blueshifted qBIC at a wavelength of $420$~nm. We choose $S_2=10$~nm to substantially decrease its total decay rate $\gamma^{(\mathrm{o})}$ [see Fig.~\ref{fig:fig3}(b)]. To benchmark the $S_2$-related sensitivity improvement, we compare the structure with $S_2=10$~nm against a symmetric one with the same fundamental amplitude ($C_1=10$~nm) but with $S_2=0$. 
According to Eq.~\eqref{eq:omega_pm}, the symmetric Fourier surface has a single bright mode at $\omega_C$ [see Eq.~\eqref{eq:omega_C}]. Therefore, we set the period of this symmetric Fourier surface to $\Lambda=251$~nm, ensuring that both structures with and without $S_2$ have  resonances at about a $420$~nm wavelength. 

Next, we calculate the reflectance spectra of these Fourier surfaces (with and without $S_2$) using FEM under different environmental refractive indices  $n_\mathrm{d}+\Delta n$, where $-0.01\le\Delta n\le0.01$ [see Fig.~\ref{fig:fig5}(b)]. 
As seen from the plots, a nonzero $S_2$ indeed reduces the overall losses, causing the resonances to become sharper and approach the critical coupling condition with $\min R\approx0$. This makes the spectral shifts induced by a nonzero $\Delta n$ more pronounced compared to the symmetric Fourier surface defined by a single cosine harmonic. The differences become more dramatic if we further increase the fundamental amplitude to $C_1=20$~nm. According to the theory, the radiative losses of the bright even mode of the symmetric structure ($S_2=0$) grow nonlinearly with $C_1$, significantly broadening the resonance. Consequently, the reflectance spectra become barely distinguishable under environmental perturbation $\Delta n$, making such a structure unfeasible for refractive index sensing [see Fig.~\ref{fig:fig5}(b)]. In contrast, due to its reduced total decay rates, the asymmetric structure with $S_2=10$~nm  still exhibits a noticeable resonance shift in response to subtle $\Delta n$.

To compare in more detail symmetric ($S_2=0$) and asymmetric ($S_2=10$~nm) Fourier surfaces, we calculate and plot the standard characteristics of their sensing performance \cite{maksimov2020refractive,yang2025permittivity} as functions of $C_1$: FWHM ($\Delta\lambda$), sensitivity $\sigma=\partial\lambda_{\mathrm{res}}/\partial n_\mathrm{d}$ calculated at $n_\mathrm{d}=1.33$, and the corresponding figure of merit $\mathrm{FOM}=\sigma/\Delta\lambda$. As seen in  Fig.~\ref{fig:fig5}(c), the FWHMs differ by more than a factor of two, and for relatively large values of $C_1$ the asymmetric structure ($S_2=10$~nm) still demonstrates feasible  linewidths, whereas the Fourier surface with no sine harmonic exhibits an excessively broad resonance, see Fig.~\ref{fig:fig5}(b). In terms of the sensitivity $\sigma$, the benefit of using an additional sine harmonic is less pronounced, yielding an improvement by only $15$--$20\%$ [see Fig.~\ref{fig:fig5}(d)]. Moreover, the sensitivity remains nearly constant with increasing $C_1$. Nevertheless, the improvements in both FWHM and $\sigma$ produced by nonzero $S_2$ combine into a considerable enhancement of the FOM in Fig.~\ref{fig:fig5}(e): across the entire considered range of $C_1$, the asymmetric structure exhibits a FOM more than twice higher than of its symmetric counterpart. Note that the absolute values of $\sigma$ or FOM are the same order of magnitude as those demonstrated by conventional optical plasmonic nanostructures \cite{ye2014high, cai2019solution, li2023high}. These values can be further optimized by reducing material absorption and adjusting structural parameters, while our present goal is strictly the relative comparison of the two geometries (with and without $S_2$) rather than global performance maximization. 
Thus, for refractive index sensing, Fourier surfaces incorporating an additional sine harmonic significantly outperform symmetric periodic profiles. 

Finally, we note that similar sensing enhancement based on even/odd eigenmode intricate interplay may be a general feature inherent to other types of surface corrugation. Fourier surfaces offer a convenient platform to achieve the optimal balance between the coupling of eigenmodes and their radiative and dissipative losses.

\phantomsection\pdfbookmark[1]{Discussion}{sec:discussion}
\section*{Discussion}
In this work, we revisited the long-standing problem of SPPs on harmonically corrugated metal surfaces and developed a first-principles eigenstate theory based on the Rayleigh hypothesis. This approach yields an explicit coupled-oscillator dispersion equation describing the evolution of the underlying eigenmodes across the shallow-to-deep grating transition and explained the corresponding far-field resonant features, previously treated separately: the PBG and the single sharp qBIC resonance.


Based on the analytical prediction that the decay rate of one of the hybrid modes decreases as $S_2$ increases, we demonstrated how to significantly improve the refractive index sensing performance. When targeting a specific sensing wavelength, it is far more effective to use an asymmetric structure (incorporating a second Fourier harmonic) rather than a single cosine grating. 
Although incorporating an additional sine harmonic might seem technically challenging, the resulting asymmetric structure requires a larger period to compensate for the $S_2$-induced blueshift and to target a specific operational wavelength (e.g., $\Lambda=290$~nm versus $\Lambda=251$~nm in our comparison). This larger period might facilitate the overall fabrication process.


We note that our theory is implemented in its simplest form, taking into account only the $-1^{\mathrm{st}}$, $0^{\mathrm{th}}$, and $+1^{\mathrm{st}}$ field harmonics. As we demonstrate in Fig.~\ref{fig:SI_RH_with_dif_Q} as well as in \cite{antonov2019corrugated}, the RH is capable of capturing and quantitatively reproducing many resonant features of Fourier surfaces with complex but smooth periodic corrugations. Therefore, our approach can be further generalized by incorporating higher-order field harmonics in order to increase quantitative accuracy or to provide deeper insights into more complicated eigensolutions. Moreover, the RH-based framework can be expanded onto in-plane momentum space \cite{antonov2021dielectric}, where oblique incidence can either play the role of the $S_2$-determined surface perturbation (see Fig.~\ref{fig:SI_R_vs_s2_and_phi}) or, in combination with this perturbation, lead to new intriguing resonant phenomena. Finally, the approach can be generalized onto 2D Fourier surfaces, where the periodic corrugation is a function of both $x$ and $y$ coordinates \cite{lassaline2020optical}.

A further promising prospect is the extension of our approach to other material platforms. In particular, the developed theory can be applied not only to metallic systems but also to polar crystals supporting surface phonon polaritons within their Reststrahlen bands, where collective lattice vibrations create spectral regions with a negative real part of the permittivity. In this case, surface plasmons are replaced by their phononic counterparts, while the general eigenstate formalism developed in this work, remains the same. More broadly, our approach can be generalized to isotropic polar dielectrics such as SiC, uniaxial hyperbolic materials such as hBN \cite{Lassaline2021,Menabde2025}, as well as to more exotic anisotropic polar crystals, such as orthorhombic $\alpha$-$\mathrm{MoO}_3$, which supports  direction-dependent phonon-polaritonic excitations.


\begin{acknowledgments}
\textbf{Acknowledgments}: Funded by the European Union (EIC, OMICSENS, 101129734, ERC, METANEXT, 101078018). Views and opinions expressed are however those of the author(s) only and do not necessarily reflect those of the European Union or the European Research Council Executive Agency. Neither the European Union nor the granting authority can be held responsible for them. This project was also funded by the Deutsche Forschungsgemeinschaft (DFG, German Research Foundation) under Germany’s Excellence Strategy EXC 3120/1 – 533771286 and EXC 2089/1–390776260 and the Emmy Noether Programme (TI 1063/1), the Bavarian program Solar Energies Go Hybrid (SolTech) and the Center for NanoScience (CeNS). The work of M.V.G. was supported by the Ministry of Science and Higher Education of the Russian Federation, project FSWU-2023-0075.
S.J.W.V. acknowledges support from the SNSF (grant no. 200021-232257). The research of D.J.N. is supported by ETH Zurich.
The authors thank Stefan A. Maier for
valuable discussions.

\noindent\textbf{Author contributions}: A.A.A., C.H., H.N., Y.M.G. and S.J.W.V. conceived the idea. A.A.A. and M.V.G. developed the theory. A.A.A. and C.H. performed numerical simulations. A.T., M.V.G., and D.J.N. supervised the project.
\end{acknowledgments}

\bibliography{Fourier}

\newcommand{\TitleFont}{\fontsize{18}{20}\selectfont\sffamily\bfseries}
\clearpage
\onecolumngrid
\makeatletter
\let\addcontentsline\orig@addcontentsline
\makeatother
\phantomsection
\pdfbookmark[0]{Supplementary Materials}{supp}  
  \centering
  {\LARGE \textbf{Supporting Information for Plasmonic Fourier Surfaces Revisited: Relating Bandgaps with Bound States in the Continuum}\par}
  \vspace{1.2cm}
  {\normalsize
  Alexander A. Antonov\textsuperscript{1,2,*},
  Connor Heimig\textsuperscript{1,2},
  Hannah Niese\textsuperscript{3},
  Yannik M. Glauser\textsuperscript{3},
  Sander J. W. Vonk\textsuperscript{3},
  David J. Norris\textsuperscript{3},
  Maxim V. Gorkunov\textsuperscript{4,5,\textdagger},
  Andreas Tittl\textsuperscript{1,2,\textdaggerdbl}\par}
  \vspace{0.8cm}
  {\small
  \textsuperscript{1}\,Institute of Photonics, Hamburg University of Technology, 21073 Hamburg, Germany\\
\textsuperscript{2}\,Chair in Hybrid Nanosystems, Nanoinstitute Munich, Faculty of Physics, Ludwig-Maximilians-Universit{\"a}t M{\"u}nchen, 80539 Munich, Germany\\
\textsuperscript{3}\,Optical Materials Engineering Laboratory, Department of Mechanical and Process Engineering, ETH Zurich, 8092 Zurich, Switzerland\\
\textsuperscript{4}\,National Research Nuclear University MEPhI, Kashirskoe shosse 31, Moscow 115409, Russia\\
\textsuperscript{5}\,Theoretical Physics and Quantum Technologies Department, National University of Science and Technology `MISIS', Moscow 119049, Russia\par}
  \vspace{0.8cm}
    \centering
  {\date{Email: $^\ast$~alexander.antonov@tuhh.de / $^\dag$~gorkunov@crys.ras.ru / $^\ddag$~andreas.tittl@tuhh.de}
  \par}
  \vspace{1.0cm}
    \raggedright
    \normalsize
    \tableofcontents
  \vfill
\setcounter{section}{0}
\renewcommand{\thesection}{S\arabic{section}}
\renewcommand{\theHsection}{S\arabic{section}}

\setcounter{figure}{0}
\renewcommand{\thefigure}{S\arabic{figure}}
\renewcommand{\theHfigure}{S\arabic{figure}}

\setcounter{table}{0}
\renewcommand{\thetable}{S\arabic{table}}
\renewcommand{\theHtable}{S\arabic{table}}

\setcounter{equation}{0}
\renewcommand{\theequation}{S\arabic{equation}}
\renewcommand{\theHequation}{S\arabic{equation}}

\section{Derivation of the dispersion equation}\label{SI:Note1}
The relation between eigenmode field harmonics in dielectric ($a_u$) and metal ($b_u$) at the interface $z=\zeta(x)$ is given by a set of linear equations (see Analytical theory section of the main text): 
\begin{equation} \label{set_of_rh_eq_SI}
\begin{gathered}
 \sum_{u}a_{u}I_{vu}  = \sum_{u} b_{u}J_{vu}\\
	-\sum_{u}a_{u}I_{vu}\frac{k_{0}^{2}-p_{u}p_{v}}{\varepsilon_\mathrm{d}k_{u}}  =
		\sum_{u}\frac{\varkappa_{0}^{2}-p_{u}p_{v}}{\varepsilon_\mathrm{m}\varkappa_{u}}b_{u}J_{vu},
        \end{gathered}
\end{equation}
where $u$ and $v$ are the integers, and the integrals $I_{vu}$ and $J_{vu}$ depend on the surface corrugation $\zeta(x)$:
\begin{equation}\label{eq:integrals_SI}
\begin{gathered}
I_{vu}=\frac{1}{\Lambda}\int_{0}^{\Lambda}\exp\left[i(u-v)\frac{2\pi}{\Lambda}x- ik_{u}\zeta(x)\right]dx,\\
	J_{vu}=\frac{1}{\Lambda}\int_{0}^{\Lambda}\exp\left[i(u-v)\frac{2\pi}{\Lambda}x+ i\varkappa_{u}\zeta(x)\right]dx.
    \end{gathered}
\end{equation}
The corresponding wavenumbers in the dielectric ($k_u$) with permittivity $\varepsilon_\mathrm{d}$ and in the metal ($\varkappa_u$) with $\varepsilon_\mathrm{m}$ are given by:
\begin{equation}
     k_u=\sqrt{\varepsilon_\mathrm{d}\left(\frac{2\pi}{\lambda}\right)^2-p^2_u}, \ \ \ \varkappa_u=\sqrt{\varepsilon_\mathrm{m}\left(\frac{2\pi}{\lambda}\right)^2-p^2_u},
\end{equation}
where $p=2\pi/\Lambda$ is the lattice wavenumber, and $p_u=pu$.

Using the system of equations~\eqref{set_of_rh_eq_SI}, the dispersion equation can be readily written in the form: 
\begin{equation}\label{eq:disp_SI}
	\det(\hat{J}_k^{-1}\hat{I}_k+\hat{J}^{-1}\hat{I})=0,
\end{equation}
where the matrices are defined as:
\begin{equation}
\label{eq:IJmatrices_SI}
\centering
\begin{array}{c}
\hat I =
\begin{bmatrix}
 I_{11} & I_{10} & I_{1-1}\\
 I_{01} & I_{00} & I_{0-1} \\
 I_{-11} & I_{-10} & I_{-1-1}
\end{bmatrix},
\quad
\hat J =
\begin{bmatrix}
 J_{11} & J_{10} & J_{1-1} \\
 J_{01} & J_{00} & J_{0-1} \\
 J_{-11} & J_{-10} & J_{-1-1}
\end{bmatrix}
\\[4ex]
\hat I_k =\dfrac{1}{\varepsilon_\mathrm{d}}
\begin{bmatrix}
 k_1 I_{11} & k_0 I_{10} & (k_0^2+p^2)k_1^{-1} I_{1-1}\\[1ex]
 k_0^2k_1^{-1} I_{01} & k_0 I_{00} & k_0^2k_1^{-1} I_{0-1} \\[1ex]
 (k_0^2+p^2)k_1^{-1} I_{-11} & k_0 I_{-10} & k_1 I_{-1-1}
\end{bmatrix}
\\[6ex]
\hspace{-2.5ex}
\hat J_k =
\dfrac{1}{\varepsilon_\mathrm{m}}
\begin{bmatrix}
 \varkappa_1 J_{11} & \varkappa_0 J_{10} & (\varkappa_0^2+p^2)\varkappa_1^{-1} J_{1-1}\\[1ex]
 \varkappa_0^2\varkappa_1^{-1} J_{01} & \varkappa_0 J_{00} & \varkappa_0^2\varkappa_1^{-1} J_{0-1} \\[1ex]
 (\varkappa_0^2+p^2)\varkappa_1^{-1} J_{-11} & \varkappa_0 J_{-10} & \varkappa_1 J_{-1-1}
\end{bmatrix}.
\end{array}
\end{equation}

In the zeroth-order approximation with $C_1=0$ and $S_2=0$, only diagonal elements in Eq.~\eqref{eq:IJmatrices_SI} remain nonzero. Substituting these matrices into the dispersion equation~\eqref{eq:disp_SI} yields the solution for a flat-metal surface:
\begin{equation}\label{eq:0th order_SI}
    (\varepsilon_\mathrm{d}+\sqrt{\varepsilon_\mathrm{m}})(\varepsilon_\mathrm{d}\varkappa_1+\varepsilon_\mathrm{m} k_1 )^2=0,
\end{equation}
where only the second factor provides two degenerate solutions, corresponding to the classical surface plasmon with
the wavenumber equal to the lattice wavenumber $p$:
\begin{equation}\label{eq:omega0_SI} 
    \omega_0=pc\sqrt{\frac{\varepsilon_\mathrm{d}+\varepsilon_\mathrm{m}}{\varepsilon_\mathrm{m}\varepsilon_\mathrm{d}}}.
\end{equation}

To move to the analytical form and avoid numerical integration in Eq.~\eqref{eq:IJmatrices_SI}, we consider the limit of shallow surface corrugation with both $C_1k_0\ll1$ and $S_2k_0\ll1$. As discussed in the main text, the specific symmetry of the surface relief $\zeta(x)=C_1\cos(2\pi x/\Lambda)+S_2\sin(4\pi x/\Lambda)$ dictates that the eigenmode spectrum must be an even function of both $C_1$ and $S_2$: inverting the sign of either $C_1$ or $S_2$ leaves the spectrum invariant. Therefore, we expand the integrals in Eq.~\eqref{eq:IJmatrices_SI} up to the second order in $C_1$ and $S_2$:
\begin{equation}\label{eq:IJapprox_SI}
\begin{gathered}
I_{11}= I_{-1-1}= 1 - \frac{1}{4}k_1^2(C_1^2 + S_2^2), \ \ \ I_{10} = I_{-10} = - \frac{i}{2}k_0C_1,\\ 
I_{1-1} = -\frac{1}{8}k_1^2C_1^2 - \frac{1}{2}k_1S_2,\ \ \ 
I_{01} =I_{0-1} = -\frac{i}{2}k_1C_1,\\
I_{00} = 1 - \frac{1}{4}k_0^2(C_1^2 + S_2^2),\ \ \  
I_{-11} = -\frac{1}{8}k_1^2 C_1^2 + \frac{1}{2}k_1 S_2\\
\\
J_{11} =J_{-1-1}= 1 - \frac{1}{4}\varkappa_1^2 \left(C_1^2 + S_2^2\right), \ \ \
J_{10} = J_{-10}= \frac{i}{2}\varkappa_0 C_1,\\ 
J_{1-1}= -\frac{1}{8}\varkappa_1^2 C_1^2 + \frac{1}{2}\varkappa_1 S_2,\ \ \  
J_{01} = J_{0-1}= \frac{i}{2} \varkappa_1 C_1,\\
J_{00} = 1 - \frac{1}{4}\varkappa_0^2\left(C_1^2 + S_2^2\right),\ \ \  
J_{-11} = -\frac{1}{8}\varkappa_1^2 C_1^2 - \frac{1}{2}\varkappa_1 S_2.
\end{gathered}
\end{equation}
Note that coefficients $J_{vu}$ can be obtained from $I_{vu}$ by the substitution $k_u\to-\varkappa_u$.

For nonzero corrugation amplitudes we evaluate all wavenumbers entering Eq.~\eqref{eq:IJapprox_SI} at the zero-order frequency~\eqref{eq:omega0_SI} and denote them by the superscript $(0)$:
\begin{equation} \label{eq:zeroapprox_SI}
\begin{gathered}
    k_0^{(0)}=p\sqrt{1+\frac{\varepsilon_\mathrm{d}}{\varepsilon_\mathrm{m}}}, \ k_1^{(0)}=ip\sqrt{-\frac{\varepsilon_\mathrm{d}}{\varepsilon_\mathrm{m}}},  \\  \varkappa_0^{(0)}=ip\sqrt{-1-\frac{\varepsilon_\mathrm{m}}{\varepsilon_\mathrm{d}}}, \ \varkappa_1^{(0)}=ip\sqrt{-\frac{\varepsilon_\mathrm{m}}{\varepsilon_\mathrm{d}}}.
    \end{gathered}
\end{equation}

Next, we assume that the eigenfrequencies $\omega_\zeta$ of the shallow corrugated Fourier surface experience a small detuning $\Delta\omega$ from the flat-metal eigenfrequency $\omega_0$:
\begin{equation}
\omega_\zeta\approx\omega_0+\Delta\omega,   
\end{equation}
and then approximately:
\begin{equation}\label{eq:k_approx_SI}
\begin{gathered}
    k_0\approx k_0^{(0)}+\frac{\Delta\omega}{c}W_\mathrm{d}\sqrt{\varepsilon_\mathrm{d}},\ \ \ k_1\approx k_1^{(0)}-\varkappa_0^{(0)}\frac{\Delta\omega}{pc}W_\mathrm{d}\sqrt{\varepsilon_\mathrm{d}}, \\
    \varkappa_0\approx \varkappa_0^{(0)}+\varkappa_1^{(0)}\frac{\Delta\omega}{pc}W_\mathrm{m}\sqrt{\varepsilon_\mathrm{d}},\ \ \    \varkappa_1\approx \varkappa_1^{(0)}+\varkappa_0^{(0)}\frac{\Delta\omega}{pc}W_\mathrm{m}\sqrt{\varepsilon_\mathrm{d}},
    \end{gathered}
\end{equation}
where the factors:
\begin{equation}
    W_j=1+\frac{\omega_0}{2\varepsilon_j(\omega_0)}\frac{d\varepsilon_j}{d\omega}(\omega_0), \ \ \ j=\mathrm{m},\mathrm{d}.
\end{equation}
account for the permittivity frequency dispersion. 

Substituting \eqref{eq:k_approx_SI} into the matrix elements \eqref{eq:IJapprox_SI}, evaluating the determinant \eqref{eq:disp_SI} and neglecting the higher order terms we arrive at the dispersion equation:
\begin{equation}\label{eq:deltaequation_SI}
\begin{gathered}
        (\Delta\omega)^2 + pc\frac{(pC_1)^2}{2} \frac{\sqrt{\varepsilon_\mathrm{d}}+i\sqrt{-\varepsilon_\mathrm{m}}} {W_\mathrm{m}\varepsilon_\mathrm{d}-W_\mathrm{d}\varepsilon_\mathrm{m}}  \Delta\omega\\+ (pc)^2(pS_2)^2\frac{(\varepsilon_\mathrm{d}-\varepsilon_\mathrm{m})^2}{(\varepsilon_\mathrm{d} + \varepsilon_\mathrm{m})(W_\mathrm{m}\varepsilon_\mathrm{d}-W_\mathrm{d}\varepsilon_\mathrm{m})^2}=0,
        \end{gathered}
\end{equation}
where we have neglected the term proportional to $S_2^2\Delta\omega $, as it is negligibly small compared to the term proportional to $S_2^2$.

\clearpage
\section{Evolution of the PBG for shallow reliefs}
\begin{figure}[H]
\centering
\includegraphics{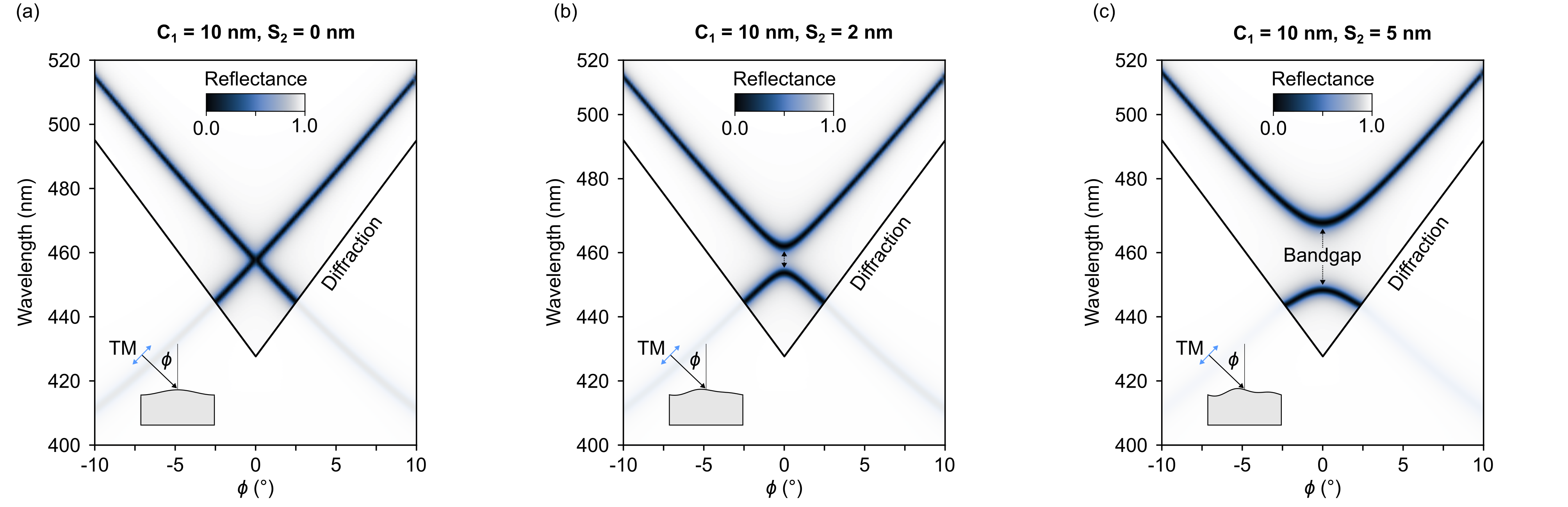}
\caption{\label{fig:Bandgap} 
Simulated reflectance spectra as a function of the incidence angle $\phi$ for silver \cite{mcpeak2015plasmonic} Fourier surfaces with shallow ($C_1=10$~nm) corrugations and (a) $S_2=0$, (b) $S_2=2$~nm, and (c) $S_2=5$~nm perturbation amplitudes. The shaded regions indicate the range with open diffraction channels.}
\end{figure}

\clearpage
\section{Radiative Q-factor as a function of the asymmetry $S_2$}
\begin{figure}[H]
\centering
\includegraphics{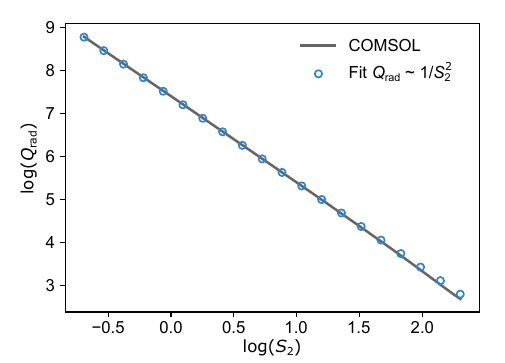}
\caption{\label{fig:SI_QvsS2} 
Radiative quality factor $Q_{\mathrm{rad}}$ as a function of the asymmetry parameter $S_2$ for a lossless Fourier surface with a deep corrugation defined ($C_1=50$~nm), a period of $\Lambda=420$~nm and $\varepsilon_\mathrm{m}=-7.2$. The log--log scale clearly reveals the characteristic qBIC scaling $Q_{\mathrm{rad}}\propto S_2^{-2}$ \cite{koshelev2018asymmetric}.
}
\end{figure}

\clearpage
\section{Symmetry breaking via Second Fourier Harmonic or in-plane momentum}
\begin{figure}[H]
\centering
\includegraphics{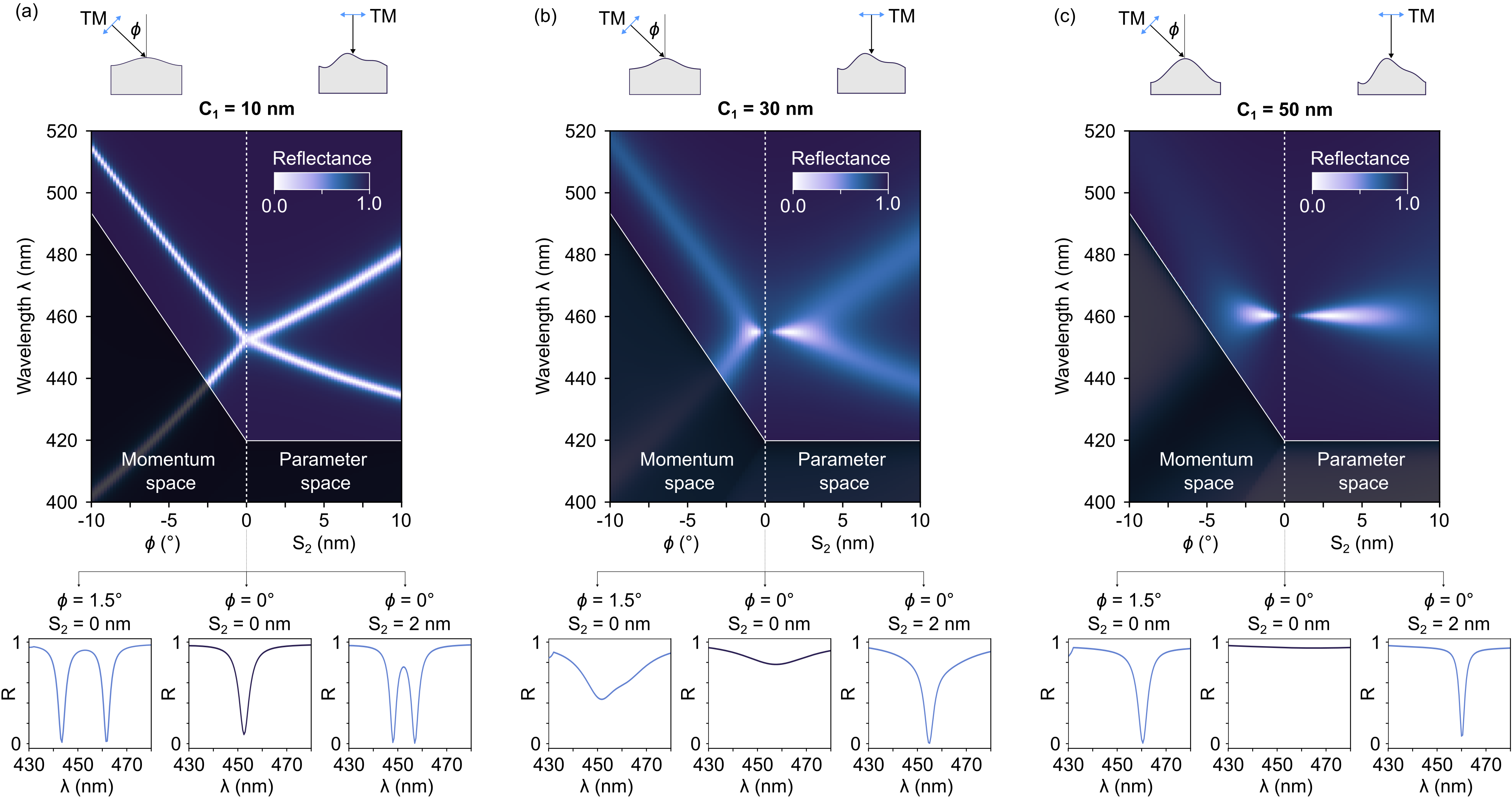}
\caption{\label{fig:SI_R_vs_s2_and_phi} 
Simulated reflectance spectra for silver \cite{mcpeak2015plasmonic} Fourier surfaces with a period of $\Lambda=420$~nm and (a) shallow ($C_1 = 10$~nm), (b) intermediate ($C_1 = 30$~nm) and (c) deep corrugations ($C_1=50$~nm). Symmetry breaking is introduced either in the parametric space via $S_2$ or in the in-plane momentum space trough the angle of incidence $\phi$. The shaded regions indicate the range with open diffraction channels.
}
\end{figure}

\clearpage
\section{Convergence of the Rayleigh hypothesis}
\begin{figure}[H]
\centering
\includegraphics{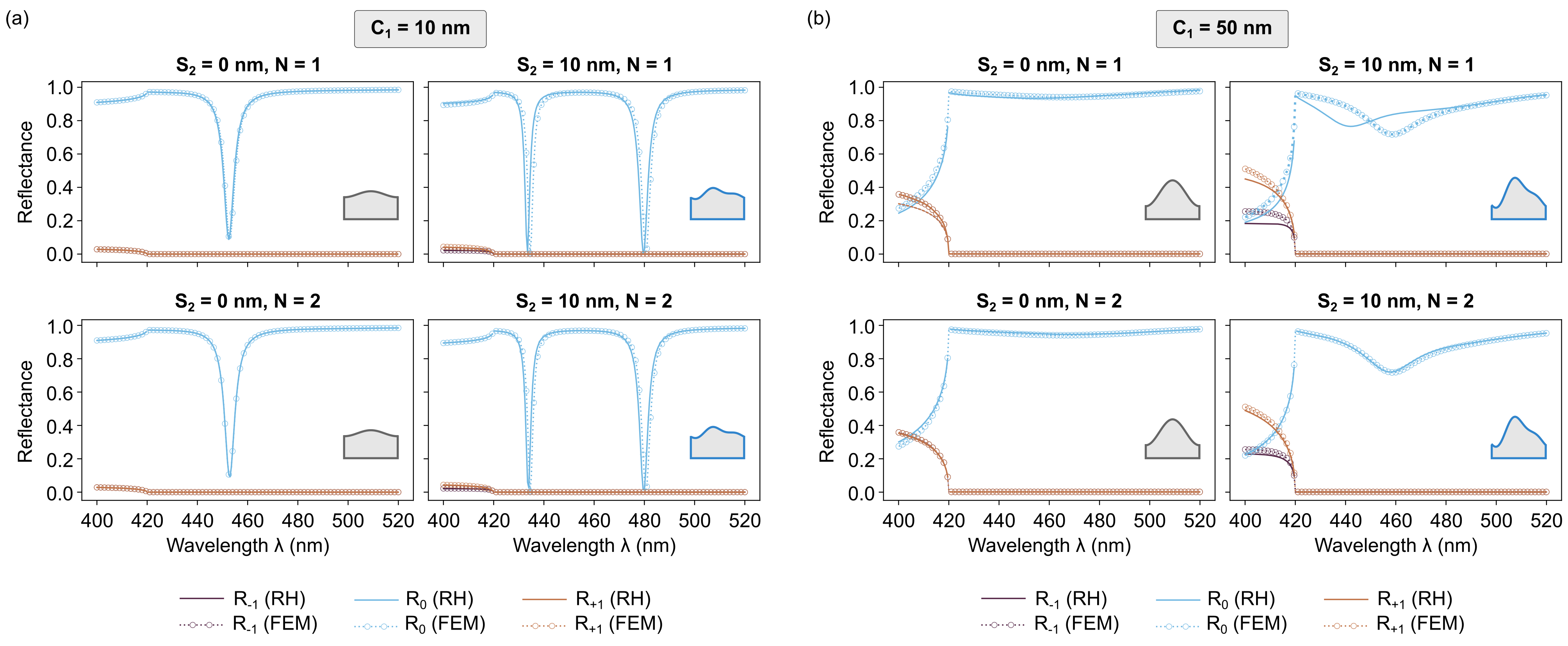}
\caption{\label{fig:SI_RH_with_dif_Q} 
Comparison between analytical theory within the RH framework (solid lines) with FEM-based simulations (dashed lines) for silver \cite{mcpeak2015plasmonic} Fourier surfaces with a period of $\Lambda=420$~nm and (a) shallow ($C_1=10$~nm) and (b) deep ($C_1=50$~nm) corrugations, respectively. The top row shows analytical zeroth reflectance $R_0$ and diffraction $R_{\pm1}$ spectra obtained by retaining only the $-1^{\mathrm{st}}$, $0^{\mathrm{th}}$ and $+1^{\mathrm{st}}$ field harmonics ($N=1$), whereas the bottom row also includes the $-2^{\mathrm{nd}}$ and $+2^{\mathrm{nd}}$ harmonics ($N=2$).
}
\end{figure}

\clearpage
\section{Eigenmode fields of deeply corrugated Fourier surfaces}
\begin{figure}[H]
\centering
\includegraphics{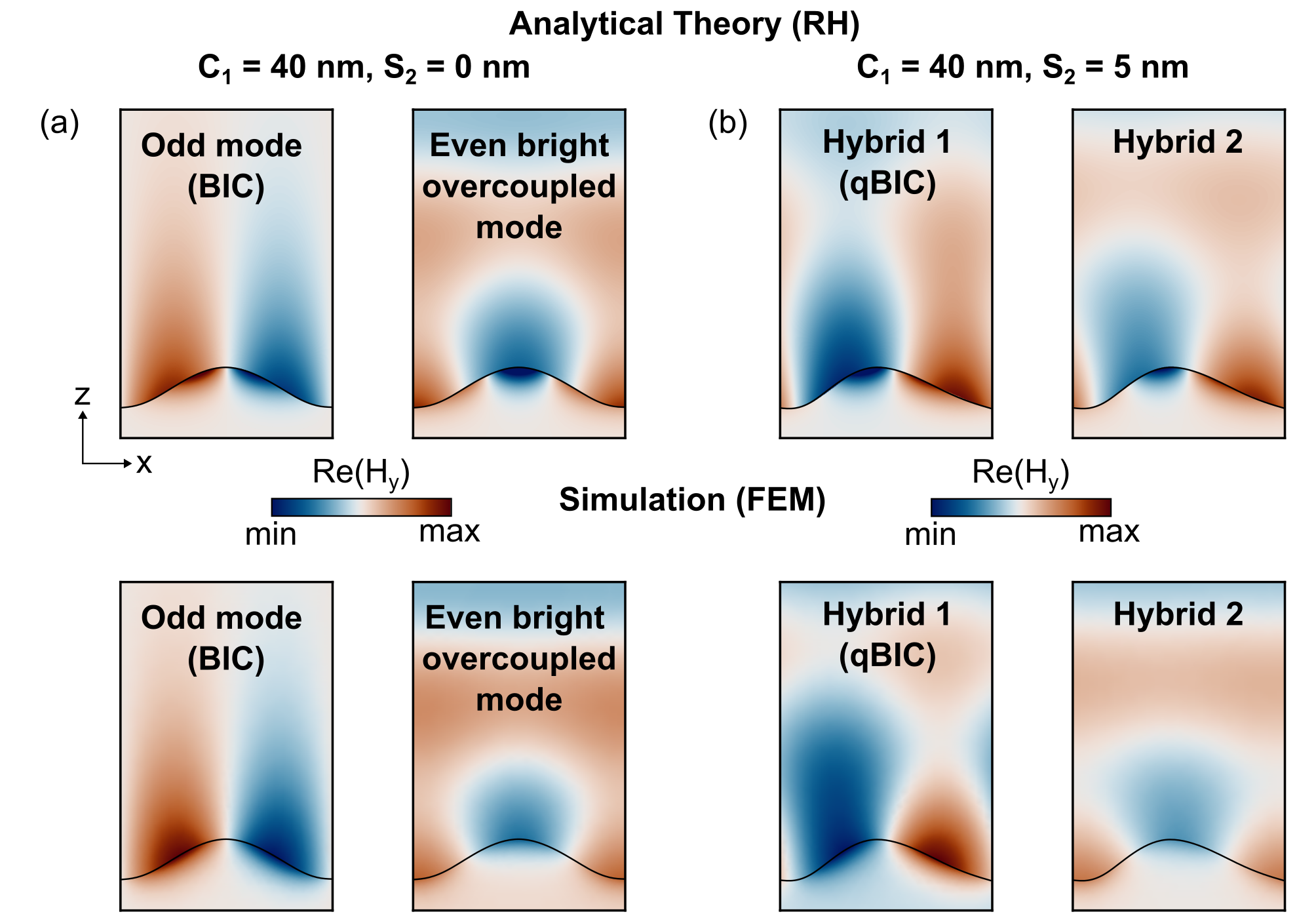 }
\caption{\label{fig:SI_Eigenfields_C1=40nm} 
Real part of the eigenmode fields $H_y(x,z)$ obtained using RH-based theory with $N=1$ (top) and FEM (bottom) for (a) symmetric ($S_2=0$~nm) and (b) asymmetric ($S_2=5$~nm) Fourier surfaces with $\varepsilon_{\mathrm{m}}=-7.2+0.2i$, a period of $\Lambda=420$~nm and a deep corrugation defined by $C_1=40$~nm.
Fields are shown over one period $\Lambda$ along the $x$-axis; the black curve denotes the metal surface.
}
\end{figure}


\end{document}